# ON THE EXACT MATRIX REPRESENTATION FOR TRANSVERSE MAGNETIC MULTIPLE SCATTERING BY AN INFINITE GRATING OF INSULATING DIELECTRIC CIRCULAR CYLINDERS AT OBLIQUE INCIDENCE

ÖMER KAVAKLIOĞLU and BARUCH SCHNEIDER
*Division of Electrophysics Research and Department of Mathematics,*
*Faculty of Computer Sciences, Izmir University of Economics,*
*Balçova, IZMIR 35330 TURKEY*
(omer_kavaklioglu@yahoo.com; omer.kavaklioglu@ieu.edu.tr; baruch.schneider@ieu.edu.tr)

## Abstract

A *'computational algorithm'* for the exact equations representing the *'scattering coefficients of an infinite grating of insulating dielectric circular cylinders associated with obliquely incident vertically polarized plane electromagnetic waves'* is generated by matrix methods, and the solution for the scattering coefficients is acquired by a matrix inversion procedure.

---

## 1. Introduction

The classical electromagnetic problem of multiple scattering of plane waves by *'periodic structures'* consisting of an infinite array made out of infinitely long insulating circular dielectric cylinders was treated by Twersky as long ago as 1956. He analyzed the problem of multiple scattering of plane waves at non-oblique arbitrary incidence on an infinite grating of circular cylinders employing the Green's function approach [1-2], and the separation-of-variables technique in terms of the functional equation approach [3].

His techniques have recently been exploited in the area of acoustical wave propagation by Cai and Williams [4-5] who treated the multiple scattering of anti-plane shear waves in fibre-reinforced composite materials in order to reconstruct the solutions for abstract scatterers. Moreover, Cai [6] investigated the *'layered multiple scattering techniques'* for anti-plane shear wave scattering from multiple gratings consisting of parallel cylinders. The more generalized case dealing with *'obliquely incident plane electromagnetic waves'* was studied by Lee [7] who demonstrated the solution for the scattering of an obliquely incident plane wave by a collection of closely-spaced, radially-stratified parallel cylinders that can have an arbitrary number of stratified layers [8].

Kavaklıoğlu [9-11] extended the results of Twersky [1-3] for the *'multiple scattering of the obliquely incident plane electromagnetic waves by an infinite grating of insulating dielectric circular cylinders'*. In a more recent investigation by Kavaklıoğlu [12], the *'direct Neumann iteration technique'* has been employed in order to acquire the exact solutions for the scattering coefficients of an infinite grating in the form of an infinite series expansion in terms of *"Wait's single scattering coefficients for an isolated dielectric circular cylinder at oblique incidence"* [13].

The purpose of this investigation is to acquire the *"exact matrix representation for the multiple scattering coefficients of an infinite grating of circular dielectric cylinders for obliquely incident vertically polarized plane electromagnetic waves"*, and capture the solution for the scattering coefficients of the infinite grating at oblique incidence. In the generalized oblique incidence solution presented in this article, the direction of the incident plane wave makes an arbitrary oblique angle of arrival $\theta_i$ with the positive *z*-axis as depicted in figure 1.

## 2. Problem formulation

### *2.1. Multiple scattering representation of the incident fields*

We consider a vertically polarized obliquely incident plane electromagnetic wave upon an infinite array of insulating circular dielectric cylinders having infinite length with radii "$a$", dielectric constant "$\varepsilon_r$", and relative permeability "$\mu_r$". The constituent cylinders of the infinite grating are placed perpendicularly to the *x-y* plane along the $y-$axis, all parallel to the $z-$axis, located at positions $\mathbf{r}_0$, $\mathbf{r}_1$, $\mathbf{r}_2$,., etc., and separated by a distance of "$d$". For this configuration, the incident wave can be written in the cylindrical coordinate system $(R_s, \phi_s, z)$ of the $s^{th}$ cylinder in terms of the cylindrical waves referred to the axis of $s^{th}$ cylinder [9-13] as

$$\mathbf{E}_v^{incident}(R_s, \phi_s, z) = \hat{\mathbf{v}}_i E_{0v} e^{ik_r sd \sin\psi_i} \left\{ \sum_{n=-\infty}^{\infty} e^{-in\psi_i} J_n(k_r R_s) e^{in(\phi_s + \pi/2)} \right\} e^{-ik_z z} \tag{1}$$

In the representation above, $\hat{\mathbf{v}}_i$ denotes the vertical polarization vector associated with a unit vector having a component parallel to all the cylinders, $\phi_i$ is the angle of incidence in *x-y* plane measured from $x-$axis in such a way that $\psi_i = \pi + \phi_i$ as it is delineated in the figure 1, implying that the wave is arbitrarily incident in the first quadrant of the coordinate system and “$J_n(x)$” denotes Bessel function of order *‘n’*. In expression (1), we have

$$k_r = k_0 \sin\theta_i \tag{2a}$$

$$k_z = k_0 \cos\theta_i \tag{2b}$$

where $k_0$ stands for the free space wavenumber, and $\theta_i$ is the *‘obliquity angle’* made with $z-$axis. “$e^{-i\omega t}$” time dependence is suppressed throughout the article, where “$\omega$” represents the angular frequency of the incident wave in radians per second and “$t$”stands for time in seconds.

*2.2. Multiple scattering representation for the z-components of the exterior fields*

The exact solution associated with the *z*-component of the electric field intensity in the exterior of the infinite grating can be expressed in terms of the incident electric field in the coordinate system of the $s^{th}$ cylinder located at $\mathbf{r}_s$, plus a summation of cylindrical waves outgoing from the individual $n^{th}$ cylinder located at $\mathbf{r}_n$, as $|\mathbf{r}-\mathbf{r}_n| \to \infty$, i. e.,

$$E_z^{(exterior)}(R_s,\phi_s,z) = E_z^{(incident)}(R_s,\phi_s,z) + \sum_{n=-\infty}^{+\infty} E_z^{(n)}(R_n,\phi_n,z) \tag{3}$$

The exterior electric and magnetic field intensities corresponding to the *‘vertically polarized obliquely incident plane electromagnetic waves’* are given in Kavaklıoğlu [9] as

$$E_z^{(exterior)}(R_s,\phi_s,z) = \left\{ e^{ik_r sd\sin\psi_i} \sum_{n=-\infty}^{+\infty} \left[ \left(E_n^i + Q_n\right) J_n(k_r R_s) + A_n H_n^{(1)}(k_r R_s) \right] e^{in(\phi_s + \pi/2)} \right\} e^{-ik_z z} \tag{4a}$$

$$H_z^{(exterior)}(R_s,\phi_s,z) = \left\{ e^{ik_r sd\sin\psi_i} \sum_{n=-\infty}^{+\infty} \left[ Q_n^H J_n(k_r R_s) + A_n^H H_n^{(1)}(k_r R_s) \right] e^{in(\phi_s + \pi/2)} \right\} e^{-ik_z z} \tag{4b}$$

The infinite set of undetermined coefficients $\{A_n, A_n^H\}_{n=-\infty}^{\infty}$ arising in (4a, b) denotes the multiple scattering coefficients for the infinite array of dielectric cylinders corresponding to the *'vertically polarized obliquely incident plane electromagnetic waves'*, $\forall n \ni n \in \mathbf{Z}$, and *"**Z**"* represents the set of all integers. $E_n^i$ in (4a, b) is given as

$$E_n^i = \sin\theta_i E_{0v} e^{-in\psi_i} \tag{5}$$

In addition, $Q_n$ and $Q_n^H$ represent the multiple scattering effects, and are expressed as a linear combination of the undetermined *'multiple scattering coefficients of the infinite grating at oblique incidence'* as

$$Q_n = \mathscr{G}_0 A_n + \sum_{\substack{m=-\infty \\ m\neq n}}^{\infty} \mathscr{G}_{n\text{-}m} A_m \tag{6a}$$

$$Q_n^H = \mathscr{G}_0 A_n + \sum_{\substack{m=-\infty \\ m\neq n}}^{\infty} \mathscr{G}_{n\text{-}m} A_m^H \tag{6b}$$

$\forall n \ni n \in \mathbf{Z}$. $\mathscr{G}_n(k_r d)$ is the generalized form of the *'Schlömilch series for obliquely incident waves'* (Twersky [1-3], Kavaklıoğlu [9-12]) explicitly given as

$$\mathscr{G}_n(k_r d) = \sum_{s=1}^{\infty} H_n^{(1)}(sk_r d)\left[e^{isk_r d\sin\psi_i}(-1)^n + e^{-isk_r d\sin\psi_i}\right] \tag{7}$$

which is convergent provided that $k_r d(1 \pm \sin\psi_i)/2\pi$ does not equal integers. In (7), "$H_n^{(1)}(x)$" denotes the Hankel function of first kind and of order *'n'*, $\forall n \ni n \in \mathbf{Z}$.

## 3. Derivation of the exact matrix system of equations for the multiple scattering coefficients of the infinite grating at oblique incidence

The equations describing the scattering coefficients for *'vertically polarized and obliquely incident plane electromagnetic waves'* are first derived in Kavaklıoğlu [9] as

$$\frac{K_n H_n^{(1)}(k_r a)}{i\omega\mu_0\beta_n(\mu_r)}[A_n + \frac{J_n(k_r a)}{H_n^{(1)}(k_r a)}(E_n^i + Q_n)] = -[A_n^H + \frac{\alpha_n(\mu_r)}{\beta_n(\mu_r)}Q_n^H] \tag{8a}$$

$$\frac{K_n H_n^{(1)}(k_r a)}{i\omega\varepsilon_0\beta_n(\varepsilon_r)}[A_n^H + \frac{J_n(k_r a)}{H_n^{(1)}(k_r a)}Q_n^H] = [A_n + \frac{\alpha_n(\varepsilon_r)}{\beta_n(\varepsilon_r)}(E_n^i + Q_n)] \tag{8b}$$

$\forall n \ni n \in \mathbf{Z}$. In the equations above, we have the following constants

$$K_n = \frac{nk_z}{a}(\frac{1}{k_1^2} - \frac{1}{k_r^2}) \tag{9}$$

$\forall n \ni n \in \mathbf{Z}$, where $k_1$ is defined as $k_1 = k_0\sqrt{\varepsilon_r\mu_r - \cos^2\theta_i}$, and

$$\alpha_n(\zeta_r) = \left[\frac{\dot{J}_n(k_r a)}{k_r} - \frac{\zeta_r}{k_1}\frac{J_n(k_r a)\,\dot{J}_n(k_1 a)}{J_n(k_1 a)}\right] \tag{10a}$$

$$\beta_n(\zeta_r) = \left[\frac{\dot{H}_n^{(1)}(k_r a)}{k_r} - \frac{\zeta_r}{k_1}\frac{H_n^{(1)}(k_r a)\,\dot{J}_n(k_1 a)}{J_n(k_1 a)}\right] \tag{10b}$$

for $\zeta \in \{\varepsilon, \mu\}$, $\forall n \ni n \in \mathbf{Z}$. $\dot{J}_n(\varsigma)$ and $\dot{H}_n^{(1)}(\varsigma)$ in equation (10a, b) are defined as the first derivatives of these functions with respect to their arguments, i.e., $\dot{J}_n(\varsigma) \equiv \frac{d}{d\varsigma}J_n(\varsigma)$ and $\dot{H}_n^{(1)}(\varsigma) \equiv \frac{d}{d\varsigma}H_n^{(1)}(\varsigma)$. Introducing some new constants for $\zeta \in \{\varepsilon, \mu\}$ and $\forall n \ni n \in \mathbf{Z}$ as

$$a_n^{\zeta} = \frac{\alpha_n(\zeta_r)}{\beta_n(\zeta_r)} \tag{11a}$$

$$b_n^{\zeta} = \frac{K_n}{i\omega\zeta_0}\frac{H_n^{(1)}(k_r a)}{\beta_n(\zeta_r)} \tag{11b}$$

$$c_n = \frac{J_n(k_r a)}{H_n^{(1)}(k_r a)} \tag{11c}$$

and employing (6a, b), we can interpret the exact form of the equations for the scattering coefficients in (8a, b)as follows:

$$b_n^{\varepsilon}[A_n^H + c_n \sum_{m=-\infty}^{+\infty} A_m^H \vartheta_{n-m}(k_r d)] = A_n + a_n^{\varepsilon}[E_n^i + \sum_{m=-\infty}^{+\infty} A_m \vartheta_{n-m}(k_r d)] \tag{12a}$$

$$b_n^{\mu}\{A_n + c_n[(E_n^i + \sum_{m=-\infty}^{+\infty} A_m \vartheta_{n-m}(k_r d)]\} = -[A_n^H + a_n^{\mu} \sum_{m=-\infty}^{+\infty} A_n^H \vartheta_{n-m}(k_r d)] \tag{12b}$$

$\forall n \ni n \in Z$. Equations (12a, b) are the equivalents of (8a, b) written in simplified notation, and the constants appearing in these equations can be determined using (11a, b, c) as

$$a_n^{\zeta} = \left[\frac{J_n(k_1 a)\,\dot{J}_n(k_r a) - \zeta_r(\frac{k_r}{k_1})\,J_n(k_r a)\,\dot{J}_n(k_1 a)}{J_n(k_1 a)\,\dot{H}_n^{(1)}(k_r a) - \zeta_r(\frac{k_r}{k_1})\,H_n^{(1)}(k_r a)\,\dot{J}_n(k_1 a)}\right] \tag{13a}$$

$$b_n^{\zeta} = \frac{K_n k_r}{i\omega\zeta_0}\left[\frac{J_n(k_1 a)H_n^{(1)}(k_r a)}{J_n(k_1 a)\dot{H}_n^{(1)}(k_r a) - \zeta_r(\frac{k_r}{k_1})H_n^{(1)}(k_r a)\dot{J}_n(k_1 a)}\right] \tag{13b}$$

for $\zeta \in \{\varepsilon, \mu\}$, and $\forall n \ni n \in \mathbf{Z}$. Utilizing (5b) and (9) together in (13b), we have deduced that

$$\frac{K_n k_r}{i\omega\zeta_0} = \frac{iF}{c\zeta_0}\left(\frac{n}{k_r a}\right) \tag{14}$$

for $\zeta \in \{\varepsilon, \mu\}$ and $\forall n \ni n \in \mathbf{Z}$, where *'c'* stands for the speed of light in free space as

$$c = \frac{1}{\sqrt{\mu_0 \varepsilon_0}} \tag{15}$$

and, *'F'* is defined as

$$F = \frac{(\mu_r \varepsilon_r - 1)\cos\theta_i}{\mu_r \varepsilon_r - \cos^2\theta_i} \tag{16}$$

Inserting (14) into (13b), we have gotten

$$b_n^{\zeta} = \frac{iF}{c\zeta_0}\left[\frac{J_n(k_1 a)H_n^{(1)}(k_r a)}{J_n(k_1 a)\dot{H}_n^{(1)}(k_r a) - \zeta_r(\frac{k_r}{k_1})H_n^{(1)}(k_r a)\dot{J}_n(k_1 a)}\right]\left(\frac{n}{k_r a}\right) \tag{17}$$

for $\zeta \in \{\varepsilon, \mu\}$, and $\forall n \ni n \in \mathbf{Z}$. Defining,

$$\xi_0 = \sqrt{\frac{\mu_0}{\varepsilon_0}} := \eta_0^{-1} \cong 377\,\Omega \tag{18}$$

which is the intrinsic impedance of free space. Employing (18) in (17), for $\zeta \in \{\varepsilon, \mu\}$, we have obtained

$$b_n^{\varepsilon} = \left[\frac{J_n(k_1 a)H_n^{(1)}(k_r a)}{J_n(k_1 a)\dot{H}_n^{(1)}(k_r a) - \varepsilon_r(\frac{k_r}{k_1})H_n^{(1)}(k_r a)\dot{J}_n(k_1 a)}\right]\left(\frac{inF\xi_0}{k_r a}\right) \tag{19a}$$

$$b_n^{\mu} = \left[ \frac{J_n(k_1 a) H_n^{(1)}(k_r a)}{J_n(k_1 a) \dot{H}_n^{(1)}(k_r a) - \mu_r (\frac{k_r}{k_1}) H_n^{(1)}(k_r a) \dot{j}_n(k_1 a)} \right] \left( \frac{inF\eta_0}{k_r a} \right) \qquad (19b)$$

$\forall n \ni n \in \mathbf{Z}$. Rearranging the equations in (12a, b), we have obtaied

$$A_n + a_n^{\varepsilon}[E_n^i + \sum_{m=-\infty}^{+\infty} A_m\, \vartheta_{n-m}(k_r d\ )] - b_n^{\varepsilon}[A_n^H + c_n \sum_{m=-\infty}^{+\infty} A_m^H\, \vartheta_{n-m}(k_r d\ )] = 0 \qquad (20a)$$

$$[A_n^H + a_n^{\mu} \sum_{m=-\infty}^{m=+\infty} A_n^H\, \vartheta_{n-m}(k_r d\ )] + b_n^{\mu}\{A_n + c_n[(E_n^i + \sum_{m=-\infty}^{+\infty} A_m\, \vartheta_{n-m}(k_r d\ )]\} = 0 \qquad (20b)$$

$\forall n \ni n \in \mathbf{Z}$. Using the forms $Q_n$ and $Q_n^H$ explicitly in (20a, b), we can modify the equations for the scattering coefficients as

$$(1 + a_n^{\varepsilon}\, \vartheta_0) A_n + a_n^{\varepsilon} \sum_{\substack{m=-\infty \\ m\neq n}}^{+\infty} A_m\, \vartheta_{n-m} - b_n^{\varepsilon}[(1 + c_n\, \vartheta_0) A_n^H + c_n \sum_{\substack{m=-\infty \\ m\neq n}}^{+\infty} A_m^H\, \vartheta_{n-m}] = -a_n^{\varepsilon} E_n^i \qquad (21a)$$

and,

$$(1 + a_n^{\mu}\, \vartheta_0) A_n^H + a_n^{\mu} \sum_{\substack{m=-\infty \\ m\neq n}}^{+\infty} A_m^H\, \vartheta_{n-m} + b_n^{\mu}[(1 + c_n\, \vartheta_0) A_n + c_n \sum_{\substack{m=-\infty \\ m\neq n}}^{+\infty} A_m\, \vartheta_{n-m}] = -b_n^{\mu} c_n E_n^i \qquad (21b)$$

$\forall n \ni n \in \mathbf{Z}$. The equations (21a, b) for the sepecial case of *'n=0'* will be simplified as

$$A_0 = -[a_0^{\varepsilon}/(1 + a_0^{\varepsilon}\, \vartheta_0)][E_0^i + \sum_{\substack{m=-\infty \\ m\neq 0}}^{+\infty} A_m\, \vartheta_{-m}] \qquad (22a)$$

$$A_0^H = -[a_0^{\mu}/(1 + a_0^{\mu}\, \vartheta_0)][\sum_{\substack{m=-\infty \\ m\neq 0}}^{+\infty} A_m^H\, \vartheta_{-m}\ ] \qquad (22b)$$

Indicating those scattering coefficients corresponding to $A_0$ and $A_0^H$ explicitly in (21a, b), we have

$$(1 + a_n^{\varepsilon}\, \vartheta_0) A_n + a_n^{\varepsilon}[A_0\, \vartheta_n + \sum_{\substack{m=-\infty \\ m\neq 0,n}}^{+\infty} A_m\, \vartheta_{n-m}]$$

$$- b_n^{\varepsilon}[(1 + c_n\, \vartheta_0) A_n^H + c_n (A_0^H\, \vartheta_n + \sum_{\substack{m=-\infty \\ m\neq 0,n}}^{+\infty} A_m^H\, \vartheta_{n-m})] = -a_n^{\varepsilon} E_n^i \qquad (23a)$$

and,

$$(1 + a_n^{\mu}\, \vartheta_0) A_n^H + a_n^{\mu}[A_0^H\, \vartheta_n + \sum_{\substack{m=-\infty \\ m\neq 0,n}}^{+\infty} A_m^H\, \vartheta_{n-m}\ ]$$

$$+ b_n^{\mu}[(1 + c_n\, \vartheta_0) A_n + c_n (A_0\, \vartheta_n + \sum_{\substack{m=-\infty \\ m\neq 0,n}}^{+\infty} A_m\, \vartheta_{n-m})] = -b_n^{\mu} c_n E_n^i \qquad (23b)$$

$\forall n \ni \{n \in Z \mid n \neq 0\}$. Similarly, for the special case with *'n=0'*, (22a, b) can be written as

$$A_0 = -[a_0^{\varepsilon}/(1+a_0^{\varepsilon}\,\vartheta_0)][E_0^{i} + (A_n\,\vartheta_{-n} + \sum_{\substack{m=-\infty\\ m\neq 0,n}}^{+\infty} A_m\,\vartheta_{-m})] \tag{24a}$$

for the scattering coefficients corresponding to the electric field, and

$$A_0^{H} = -[a_0^{\mu}/(1+a_0^{\mu}\,\vartheta_0)][A_n^{H}\,\vartheta_{-n} + \sum_{\substack{m=-\infty\\ m\neq 0,n}}^{+\infty} A_m^{H}\,\vartheta_{-m}] \tag{24b}$$

for the scattering coefficients corresponding to the magnetic field. We have noticed that the equations (24a, b) represent the solutions for $A_0$, and $A_0^H$ which are expressed in terms of all the other scattering cofficients $\{A_n, A_n^H\}$, $\forall n \ni \{n \in Z \mid n \neq 0\}$. Therefore, we can insert these solutions of $A_0$ and $A_0^H$ into (23a, b) to eliminate $A_0$ and $A_0^H$ terms from the equations of the scattering coefficients. For this purpose, we have first used (24a, b) in the evaluation of the following identities as

$$a_n^{\varepsilon}[A_0\,\vartheta_n + \sum_{\substack{m=-\infty\\ m\neq 0,n}}^{+\infty} A_m\,\vartheta_{n-m}] \equiv -[a_0^{\varepsilon}/(1+a_0^{\varepsilon}\,\vartheta_0)]\,a_n^{\varepsilon}\,\vartheta_n E_0^{i} - [a_0^{\varepsilon}/(1+a_0^{\varepsilon}\,\vartheta_0)]\,\vartheta_n\vartheta_n a_n^{\varepsilon} A_n$$

$$+a_n^{\varepsilon} \sum_{\substack{m=-\infty\\ m\neq 0,n}}^{+\infty} A_m\{\vartheta_{n-m} - [a_0^{\varepsilon}/(1+a_0^{\varepsilon}\,\vartheta_0)]\,\vartheta_n\vartheta_{-m}\} \tag{25a}$$

$\forall n \ni \{n \in Z \mid n \neq 0\}$, and

$$a_n^{\mu}[A_0^{H}\,\vartheta_n + \sum_{\substack{m=-\infty\\ m\neq 0,n}}^{+\infty} A_m^{H}\,\vartheta_{n-m}] \equiv -[a_0^{\mu}/(1+a_0^{\mu}\,\vartheta_0)]\,\vartheta_n\vartheta_{-n} a_n^{\mu} A_n^{H}$$

$$+a_n^{\mu} \sum_{\substack{m=-\infty\\ m\neq 0,n}}^{+\infty} A_m^{H}\{\vartheta_{n-m} - [a_0^{\mu}/(1+a_0^{\mu}\,\vartheta_0)]\,\vartheta_n\vartheta_{-m}\} \tag{25b}$$

$\forall n \ni \{n \in Z \mid n \neq 0\}$. Using (25a, b) in (23a, b), we have gotten the equations for the scattering coefficients of the infinite grating of circular dielectric cylinders as

$$\{1 + a_n^{\varepsilon}\{\vartheta_0 - [a_0^{\varepsilon}/(1+a_0^{\varepsilon}\,\vartheta_0)]\,\vartheta_n\vartheta_{-n}\}\}\,A_n$$

$$+a_n^{\varepsilon} \sum_{\substack{m=-\infty\\ m\neq 0,n}}^{+\infty} A_m\{\vartheta_{n-m} - [a_0^{\varepsilon}/(1+a_0^{\varepsilon}\,\vartheta_0)]\,\vartheta_n\vartheta_{-m}\}$$

$$-\,b_n^{\varepsilon}\{A_n^{H}[1 + c_n\{\vartheta_0 - [a_0^{\mu}/(1+a_0^{\mu}\,\vartheta_0)]\,\vartheta_n\vartheta_{-n}\}]$$

$$+c_n \sum_{\substack{m=-\infty\\ m\neq 0}}^{+\infty} A_m^{H}\{\vartheta_{n-m} - [a_0^{\mu}/(1+a_0^{\mu}\,\vartheta_0)]\,\vartheta_n\vartheta_{-m}\}\}$$

$$= -a_n^{\varepsilon}\{E_n^{i} - [a_0^{\varepsilon}/(1+a_0^{\varepsilon}\,\vartheta_0)]\,\vartheta_n E_0^{i}\} \tag{26a}$$

and

$$\{1+a_n^{\mu}\{\vartheta_0-[a_0^{\mu}/(1+a_0^{\mu}\vartheta_0)]\vartheta_n\vartheta_{-n}\}\}A_n^H$$

$$+a_n^{\mu}\sum_{\substack{m=-\infty\\ m\neq 0,n}}^{+\infty}A_m^H\{\vartheta_{n-m}-[a_0^{\mu}/(1+a_0^{\mu}\vartheta_0)]\vartheta_n\vartheta_{-m}\}$$

$$+b_n^{\mu}\{A_n[1+c_n\{\vartheta_0-[a_0^{\varepsilon}/(1+a_0^{\varepsilon}\vartheta_0)]\vartheta_n\vartheta_{-n}\}]$$

$$+c_n\sum_{\substack{m=-\infty\\ m\neq 0}}^{+\infty}A_m\{-[a_0^{\varepsilon}/(1+a_0^{\varepsilon}\vartheta_0)]\vartheta_n\vartheta_{-m}\}\}=-b_n^{\mu}c_nE_n^i \quad (26b)$$

$\forall n\ni\{n\in Z\mid n\neq 0\}$. The equations in (26a, b) together represent the complete set of equations for the scattering coefficients of the infinite grating of circular dielectric cylinders at oblique incidence. Incorporating the first terms inside the brackets into the infinite summation, (26a, b) can be written more compactly as

$$A_n+a_n^{\varepsilon}\sum_{\substack{m=-\infty\\ m\neq 0}}^{+\infty}A_m\{\vartheta_{n-m}-[a_0^{\varepsilon}/(1+a_0^{\varepsilon}\vartheta_0)]\vartheta_n\vartheta_{-m}\}$$

$$-b_n^{\varepsilon}\{A_n^H+c_n\sum_{\substack{m=-\infty\\ m\neq 0}}^{+\infty}A_m^H\{\vartheta_{n-m}-[a_0^{\mu}/(1+a_0^{\mu}\vartheta_0)]\vartheta_n\vartheta_{-m}\}$$

$$=-a_n^{\varepsilon}\{E_n^i-[a_0^{\varepsilon}/(1+a_0^{\varepsilon}\vartheta_0)]\vartheta_nE_0^i\} \quad (27a)$$

and

$$A_n^H+a_n^{\mu}\sum_{\substack{m=-\infty\\ m\neq 0}}^{+\infty}A_m^H\{\vartheta_{n-m}-[a_0^{\mu}/(1+a_0^{\mu}\vartheta_0)]\vartheta_n\vartheta_{-m}\}$$

$$+b_n^{\mu}\{A_n+c_n\sum_{\substack{m=-\infty\\ m\neq 0}}^{+\infty}A_m\{\vartheta_{n-m}-[a_0^{\varepsilon}/(1+a_0^{\varepsilon}\vartheta_0)]\vartheta_n\vartheta_{-m}\}=-b_n^{\mu}c_nE_n^i \quad (27b)$$

$\forall n\ni\{n\in Z\mid n\neq 0\}$. The equations in (27a, b) represent the complete set of equations for the scattering coefficients $A_n$ and $A_n^H$'s of the infinite grating at oblique incidence. On the other hand, $A_0$ and $A_0^H$'s are obtained from (22a, b) in terms of all the other scattering coefficients, i.e., $A_n$ and $A_n^H$'s $\forall n\ni\{n\in Z\mid n\neq 0\}$.

Defining a new set of parameters $d_{n,m}^{\zeta}$, $e_n^{\zeta}$, and $f_n^{\zeta}$ for $\zeta\in\{\varepsilon,\mu\}$, and $\forall n\ni\{n\in Z\mid n\neq 0\}$ and $\forall m\ni\{m\in Z\mid m\neq 0\}$ as

$$d_{n,m}^{\zeta}=\{\vartheta_{n-m}-[a_0^{\zeta}/(1+a_0^{\zeta}\vartheta_0)]\vartheta_n\vartheta_{-m}\} \quad (28a)$$

$$e_n^{\zeta}=-a_n^{\zeta}\{E_n^i-[a_0^{\zeta}/(1+a_0^{\zeta}\vartheta_0)]\vartheta_nE_0^i\} \quad (28b)$$

$$f_n^{\zeta} = -b_n^{\zeta} c_n E_n^i \tag{28c}$$

and employing the expressions (28a, b, c) in the equations (27a, b), the equations for the *'scattering coefficients of the infinite grating associated with obliquely incident and vertically polarized waves'* can more compactly be expressed as

$$[A_n + a_n^{\varepsilon} \sum_{\substack{m=-\infty \\ m\neq 0}}^{+\infty} d_{n,m}^{\varepsilon} A_m] - b_n^{\varepsilon}[A_n^H + c_n \sum_{\substack{m=-\infty \\ m\neq 0}}^{+\infty} d_{n,m}^{\mu} A_m^H] = e_n^{\varepsilon} \tag{29a}$$

$$[A_n^H + a_n^{\mu} \sum_{\substack{m=-\infty \\ m\neq 0}}^{+\infty} d_{n,m}^{\mu} A_m^H] + b_n^{\mu}[A_n + c_n \sum_{\substack{m=-\infty \\ m\neq 0}}^{+\infty} d_{n,m}^{\varepsilon} A_m] = f_n^{\mu} \tag{29b}$$

$\forall n \ni \{n \in Z \mid n \neq 0\}$. We have obtained the following matrix system of equations from (29a, b) as

$$\{\underline{\underline{I}} + [\ Diag\ a_n^{\varepsilon}].\underline{\underline{D}}^{\varepsilon}\}.\underline{A} - [\ Diag\ b_n^{\varepsilon}].\{\underline{\underline{I}} + [\ Diag\ c_n].\underline{\underline{D}}^{\mu}\}.\underline{A}^H = \underline{e}^{\varepsilon} \tag{30a}$$

and,

$$\{\underline{\underline{I}} + [\ Diag\ a_n^{\mu}].\underline{\underline{D}}^{\mu}\}.\underline{A}^H + [\ Diag\ b_n^{\mu}].\{\underline{\underline{I}} + [\ Diag\ c_n].\underline{\underline{D}}^{\varepsilon}\}.\underline{A} = \underline{f}^{\mu} \tag{30b}$$

where, for $\zeta \in \{\varepsilon, \mu\}$. In (30a, b), we have defined

$$\left[Diag\ a_n^{\zeta}\right] = \begin{bmatrix} \cdot & & & & & & & & \\ & \cdot & & & & & \mathbf{0} & & \\ & & a_3^{\zeta} & & & & & & \\ & & & a_2^{\zeta} & & & & & \\ & & & & a_1^{\zeta} & & & & \\ & & & & & a_{-1}^{\zeta} & & & \\ & & & & & & a_{-2}^{\zeta} & & \\ & \mathbf{0} & & & & & & a_{-3}^{\zeta} & \\ & & & & & & & & \cdot \\ & & & & & & & & & \cdot \end{bmatrix} \equiv \underline{\underline{\Lambda}}^{\zeta} \tag{31}$$

where, $\underline{\underline{\Lambda}}^{\zeta}$ is an $\infty \times \infty$ diagonal matrix, defined for $\zeta \in \{\varepsilon, \mu\}$, and its elements, $a_n^{\zeta}$, are given in (13a);

$$\left[Diag\ b_n^{\zeta}\right] = \begin{bmatrix} \cdot & & & & & & & & \\ & \cdot & & & & \mathbf{0} & & & \\ & & b_3^{\zeta} & & & & & & \\ & & & b_2^{\zeta} & & & & & \\ & & & & b_1^{\zeta} & & & & \\ & & & & & b_{-1}^{\zeta} & & & \\ & & & & & & b_{-2}^{\zeta} & & \\ & \mathbf{0} & & & & & & b_{-3}^{\zeta} & \\ & & & & & & & & \cdot \\ & & & & & & & & & \cdot \end{bmatrix} \equiv \underline{\underline{\mathrm{B}}}^{\zeta} \tag{32}$$

where, $\underline{\underline{\mathrm{B}}}^{\zeta}$ is an $\infty \times \infty$ diagonal matrix, defined for $\zeta \in \{\varepsilon, \mu\}$ ; and its elements, $b_n^{\zeta}$, are given in (13b);

$$\left[Diag\ c_n\right] = \begin{bmatrix} \cdot & & & & & & & & \\ & \cdot & & & & \mathbf{0} & & & \\ & & c_3 & & & & & & \\ & & & c_2 & & & & & \\ & & & & c_1 & & & & \\ & & & & & c_{-1} & & & \\ & & & & & & c_{-2} & & \\ & \mathbf{0} & & & & & & c_{-3} & \\ & & & & & & & & \cdot \\ & & & & & & & & & \cdot \end{bmatrix} \equiv \underline{\underline{\Gamma}} \tag{33}$$

where, $\underline{\underline{\Gamma}}$ is an $\infty \times \infty$ diagonal matrix, and its elements, $c_n$, are given in (11c); and

$$\underline{\underline{\mathrm{D}}}^{\zeta} \equiv \left[d_{n,m}^{\zeta}\right] = \begin{bmatrix} \cdot & \cdot & \cdot & \cdot & \cdot & \cdot & \cdot & \cdot \\ \cdot & \cdot & \cdot & \cdot & \cdot & \cdot & \cdot & \cdot \\ \cdot & \cdot & d_{2,2}^{\zeta} & d_{2,1}^{\zeta} & d_{2,-1}^{\zeta} & d_{2,-2}^{\zeta} & \cdot & \cdot \\ \cdot & \cdot & d_{1,2}^{\zeta} & d_{1,1}^{\zeta} & d_{1,-1}^{\zeta} & d_{1,-2}^{\zeta} & \cdot & \cdot \\ \cdot & \cdot & d_{-1,2}^{\zeta} & d_{-1,1}^{\zeta} & d_{-1,-1}^{\zeta} & d_{-1,-2}^{\zeta} & \cdot & \cdot \\ \cdot & \cdot & d_{-2,2}^{\zeta} & d_{-2,1}^{\zeta} & d_{-2,-1}^{\zeta} & d_{-2,-2}^{\zeta} & \cdot & \cdot \\ \cdot & \cdot & \cdot & \cdot & \cdot & \cdot & \cdot & \cdot \\ \cdot & \cdot & \cdot & \cdot & \cdot & \cdot & \cdot & \cdot \end{bmatrix} \tag{34}$$

where, $\underline{\underline{\mathrm{D}}}^{\zeta}$ is an $\infty \times \infty$ matrix, defined for $\zeta \in \{\varepsilon, \mu\}$ , and its elements, $d_{n,m}^{\zeta}$, are given in (28a); and $\underline{\underline{\mathrm{I}}}$ is an $\infty \times \infty$ identity matrix. In addition, we have

$$\underline{e}^{\zeta} \equiv \begin{pmatrix} \cdot \\ \cdot \\ e_3^{\zeta} \\ e_2^{\zeta} \\ e_1^{\zeta} \\ e_{-1}^{\zeta} \\ e_{-2}^{\zeta} \\ e_{-3}^{\zeta} \\ \cdot \\ \cdot \end{pmatrix} \qquad (35a) \qquad \underline{f}^{\zeta} \equiv \begin{pmatrix} \cdot \\ \cdot \\ f_3^{\zeta} \\ f_2^{\zeta} \\ f_1^{\zeta} \\ f_{-1}^{\zeta} \\ f_{-2}^{\zeta} \\ f_{-3}^{\zeta} \\ \cdot \\ \cdot \end{pmatrix} \qquad (35b)$$

where, $\underline{e}^{\zeta}$ and $\underline{f}^{\zeta}$ are $\infty \times 1$ vectors, defined for $\zeta \in \{\varepsilon, \mu\}$ , and their elements $e_n^{\zeta}$, and $f_n^{\zeta}$ are given in (28b, c) respectively. On the other hand, the unknown scattering coefficients associated with the exterior electric and magnetic fields are defined as

$$\underline{\mathrm{A}} \equiv \begin{pmatrix} \cdot \\ \cdot \\ A_3 \\ A_2 \\ A_1 \\ A_{-1} \\ A_{-2} \\ A_{-3} \\ \cdot \\ \cdot \end{pmatrix} \qquad (36a) \qquad \underline{\mathrm{A}}^{H} \equiv \begin{pmatrix} \cdot \\ \cdot \\ A_3^{H} \\ A_2^{H} \\ A_1^{H} \\ A_{-1}^{H} \\ A_{-2}^{H} \\ A_{-3}^{H} \\ \cdot \\ \cdot \end{pmatrix} \qquad (36b)$$

$\underline{\mathrm{A}}$ and $\underline{\mathrm{A}}^{H}$ are $\infty \times 1$ unknown vectors for the *'scattering coefficients of the electric and magnetic fields of the infinite grating of dielectric circular cylinders corresponding to the the vertically polarized obliquely incident plane electromagnetic waves'*.

We can use the definitions of (31-36a, b) in combining (30a, b) into a single matrix system of equations for the scattering coefficients of the infinite grating at oblique incidence, and write (30a, b) as a *"unique"* matrix system of equation for the all of the scattering coefficients as

$$\begin{bmatrix} \left(\underline{\underline{I}} + \underline{\underline{\Lambda}}^{\varepsilon}.\underline{\underline{D}}^{\varepsilon}\right) & | & -\underline{\underline{B}}^{\varepsilon}.\left(\underline{\underline{I}} + \underline{\underline{\Gamma}}\ .\underline{\underline{D}}^{\mu}\right) \\ ---------- & | & ---------- \\ \underline{\underline{B}}^{\mu}.\left(\underline{\underline{I}} + \underline{\underline{\Gamma}}\ .\underline{\underline{D}}^{\varepsilon}\right) & | & \left(\underline{\underline{I}} + \underline{\underline{\Lambda}}^{\mu}.\underline{\underline{D}}^{\mu}\right) \end{bmatrix} \begin{pmatrix} \underline{A} \\ -- \\ \underline{A}^{H} \end{pmatrix} = \begin{pmatrix} \underline{e}^{\varepsilon} \\ -- \\ \underline{f}^{\mu} \end{pmatrix} \tag{37}$$

**4. The solution to the exact matrix system of equations for the scattering coefficients of the infinite grating at oblique incidence**

The '*matrix system of equations for the scattering coefficients of the exterior electric and magnetic fields corresponding to the vertically polarized obliquely incident plane electromagnetic waves*' expressed by (37) can be rewritten as two separete matrix equations as

$$\left(\underline{\underline{I}} + \underline{\underline{\Lambda}}^{\varepsilon}.\underline{\underline{D}}^{\varepsilon}\right).\underline{A} - \underline{\underline{B}}^{\varepsilon}.\left(\underline{\underline{I}} + \underline{\underline{\Gamma}}\ .\underline{\underline{D}}^{\mu}\right).\underline{A}^{H} = \underline{e}^{\varepsilon} \tag{38a}$$

$$\underline{\underline{B}}^{\mu}.\left(\underline{\underline{I}} + \underline{\underline{\Gamma}}\ .\underline{\underline{D}}^{\varepsilon}\right).\underline{A} + \left(\underline{\underline{I}} + \underline{\underline{\Lambda}}^{\mu}.\underline{\underline{D}}^{\mu}\right).\underline{A}^{H} = \underline{f}^{\mu} \tag{38b}$$

From (38a) we can solve for $\underline{A}$ as

$$\underline{A} = \left(\underline{\underline{I}} + \underline{\underline{\Lambda}}^{\varepsilon}.\underline{\underline{D}}^{\varepsilon}\right)^{-1}\left[\underline{e}^{\varepsilon} + \underline{\underline{B}}^{\varepsilon}.(\underline{\underline{I}} + \underline{\underline{\Gamma}}\ .\underline{\underline{D}}^{\mu}).\underline{A}^{H}\right] \tag{39a}$$

and, from (38b) we can solve for $\underline{A}^{H}$ as

$$\underline{A}^{H} = \left(\underline{\underline{I}} + \underline{\underline{\Lambda}}^{\mu}.\underline{\underline{D}}^{\mu}\right)^{-1}\left[\underline{f}^{\mu} - \underline{\underline{B}}^{\mu}.(\underline{\underline{I}} + \underline{\underline{\Gamma}}\ .\underline{\underline{D}}^{\varepsilon}).\underline{A}\right] \tag{39b}$$

Using the expression of (39b) for $\underline{A}^{H}$ in (39a), we have obtained the solution for $\underline{A}$ as

$$\underline{A} = \left[\underline{\underline{I}} + \left(\underline{\underline{I}} + \underline{\underline{\Lambda}}^{\varepsilon}.\underline{\underline{D}}^{\varepsilon}\right)^{-1}.\underline{\underline{B}}^{\varepsilon}.\left(\underline{\underline{I}} + \underline{\underline{\Gamma}}\ .\underline{\underline{D}}^{\mu}\right).\left(\underline{\underline{I}} + \underline{\underline{\Lambda}}^{\mu}.\underline{\underline{D}}^{\mu}\right)^{-1}.\underline{\underline{B}}^{\mu}.\left(\underline{\underline{I}} + \underline{\underline{\Gamma}}\ .\underline{\underline{D}}^{\varepsilon}\right)\right]^{-1}$$

$$.\left(\underline{\underline{I}} + \underline{\underline{\Lambda}}^{\varepsilon}.\underline{\underline{D}}^{\varepsilon}\right)^{-1}.\left[\underline{e}^{\varepsilon} + \underline{\underline{B}}^{\varepsilon}.\left(\underline{\underline{I}} + \underline{\underline{\Gamma}}\ .\underline{\underline{D}}^{\mu}\right).\left(\underline{\underline{I}} + \underline{\underline{\Lambda}}^{\mu}.\underline{\underline{D}}^{\mu}\right)^{-1}.\underline{f}^{\mu}\right] \tag{40}$$

Defining two $\infty \times \infty$ matrices in terms of the previously defined matrices as

$$\underline{\underline{\Omega}}_{\varepsilon\mu} \equiv \underline{\underline{B}}^{\varepsilon}.(\underline{\underline{I}} + \underline{\underline{\Gamma}}\ .\underline{\underline{D}}^{\mu}).(\underline{\underline{I}} + \underline{\underline{\Lambda}}^{\mu}.\underline{\underline{D}}^{\mu})^{-1} \tag{41}$$

and interchanging the places of $\varepsilon$ and $\mu$ in (41), we have defined

$$\underline{\underline{\Omega}}_{\mu\varepsilon} \equiv \underline{\underline{B}}^{\mu}.(\underline{\underline{I}} + \underline{\underline{\Gamma}} .\underline{\underline{D}}^{\varepsilon}).(\underline{\underline{I}} + \underline{\underline{\Lambda}}^{\varepsilon}.\underline{\underline{D}}^{\varepsilon})^{-1} \tag{42}$$

W can put the expression for $\underline{A}$ in (40) into

$$\underline{A} = (\underline{\underline{I}} + \underline{\underline{\Lambda}}^{\varepsilon}.\underline{\underline{D}}^{\varepsilon})^{-1} .[(\underline{\underline{I}} + \underline{\underline{\Omega}}_{\mu\varepsilon}^{-1} \cdot \underline{\underline{\Omega}}_{\varepsilon\mu}^{-1}).\underline{e}^{\varepsilon} + (\underline{\underline{\Omega}}_{\varepsilon\mu} + \underline{\underline{\Omega}}_{\mu\varepsilon}^{-1}).\underline{f}^{\mu}] \tag{43}$$

inserting the expression of $\underline{A}$ in (43) into $\underline{A}^{H}$ in (39b), we have finally obtained

$$\underline{A}^{H} = -(\underline{\underline{I}} + \underline{\underline{\Lambda}}^{\mu}.\underline{\underline{D}}^{\mu})^{-1} \cdot \left[\left(\underline{\underline{\Omega}}_{\mu\varepsilon} + \underline{\underline{\Omega}}_{\varepsilon\mu}^{-1}\right).\underline{e}^{\varepsilon} + \left(\underline{\underline{\Omega}}_{\mu\varepsilon} \cdot \underline{\underline{\Omega}}_{\varepsilon\mu}\right).\underline{f}^{\mu}\right] \tag{44}$$

Combining (43-44) into a single matrix expression, we can write

$$\begin{pmatrix} \underline{A} \\ -- \\ \underline{A}^{H} \end{pmatrix} = \left[\begin{array}{c|c} \left(\underline{\underline{I}} + \underline{\underline{\Lambda}}^{\varepsilon}.\underline{\underline{D}}^{\varepsilon}\right)^{-1} & \underline{\underline{0}} \\ \hline \underline{\underline{0}} & -\left(\underline{\underline{I}} + \underline{\underline{\Lambda}}^{\mu}.\underline{\underline{D}}^{\mu}\right)^{-1} \end{array}\right]$$

$$\times \left[\begin{array}{c|c} \left(\underline{\underline{I}} + \underline{\underline{\Omega}}_{\mu\varepsilon}^{-1}.\underline{\underline{\Omega}}_{\varepsilon\mu}^{-1}\right) & \left(\underline{\underline{\Omega}}_{\varepsilon\mu} + \underline{\underline{\Omega}}_{\mu\varepsilon}^{-1}\right) \\ \hline \left(\underline{\underline{\Omega}}_{\varepsilon\mu}^{-1} + \underline{\underline{\Omega}}_{\mu\varepsilon}\right) & \left(\underline{\underline{\Omega}}_{\mu\varepsilon}.\underline{\underline{\Omega}}_{\varepsilon\mu}\right) \end{array}\right] \begin{pmatrix} \underline{e}^{\varepsilon} \\ -- \\ \underline{f}^{\mu} \end{pmatrix} \tag{45}$$

which is the complete solution vector for the whole scattering coefficients of the infinite grating at oblique incidence.

## 5. Conclusion

In this investigation, we have derived the *'exact matrix system equations for the multiple scattering coefficients of an infinite grating associated with vertically polarized obliquely incident plane electromagnetic waves'*. In addition, we have acquired the solution for the scattering coefficients at oblique incidence by matrix inversion procedure.

## References

[1] Twersky V 1956 On the scattering of waves by an infinite grating *IRE Trans. Antennas Propagat.* **AP-4** 330

[2] Twersky V 1961 Elementary function representations of Schlömilch series *Arch. Ration. Mech. Anal.* **8** 323

[3] Twersky V 1962 On scattering of waves by the infinite grating of circular cylinders *IRE Trans. on Antennas Propagat.* **AP-10** 737

[4] Cai L-W and Williams Jr J H 1999a Large-scale multiple scattering problems *Ultrason.* **37**(7) 453

[5] Cai L-W and Williams Jr J H 1999b Full-scale simulations of elastic wave scattering in fiber reinforced composites *Ultrason.* **37**(7) 463

[6] Cai L-W 2006 Evaluation of layered multiple-scattering method for antiplane shear wave scattering from gratings *J. Acoust. Soc. Am.* **120** 49

[7] Lee S C 1990 Dependent scattering of an obliquely incident plane wave by a collection of parallel cylinders *J. Appl. Phys.* **68** 4952

[8] Lee S C 1992 Scattering by closely-spaced radially-stratified parallel cylinders. *J. Quant. Spectrosc. Radiat. Transfer* **48** 119

[9] Kavaklıoğlu Ö 2000 Scattering of a plane wave by an infinite grating of circular dielectric cylinders at oblique incidence: E-polarization *Int. J. Electron.* **87** 315

[10] Kavaklıoğlu Ö 2001 On diffraction of waves by the infinite grating of circular dielectric cylinders at oblique incidence: Floquet representation *J. Mod. Opt.* **48** 125

[11] Kavaklıoğlu Ö 2002 On Schlömilch series representation for the transverse electric multiple scattering by an infinite grating of insulating dielectric circular cylinders at oblique incidence *J. Phys. A: Math. Gen.* **35** 2229

[12] Kavaklıoğlu Ö 2007 On multiple scattering of radiation by an infinite grating of dielectric circular cylinders at oblique incidence (in review)

[13] Wait J R 1955 Scattering of a plane wave from a circular dielectric cylinder at oblique incidence *Can. J. Phys.* **33** 189